\documentclass[acmlarge,screen]{acmart}
\usepackage{graphicx}
\usepackage{tikz}
\usetikzlibrary{trees}
\usepackage{subfig}
\AtBeginDocument{%
  \providecommand\BibTeX{{%
    \normalfont B\kern-0.5em{\scshape i\kern-0.25em b}\kern-0.8em\TeX}}}

\setcopyright{acmcopyright}
\copyrightyear{2018}
\acmYear{2018}
\acmDOI{XXXXXXX.XXXXXXX}

\acmConference[Conference acronym 'XX]{Make sure to enter the correct
  conference title from your rights confirmation emai}{June 03--05,
  2018}{Woodstock, NY}
\acmPrice{15.00}
\acmISBN{978-1-4503-XXXX-X/18/06}

\begin{document}

\title[The Real Her?]{The Real Her? Exploring Whether Young Adults Accept Human-AI Love}

\author{Shuning Zhang}
\orcid{0000-0002-4145-117X}
\authornotemark[1]
\email{zsn23@mails.tsinghua.edu.cn}
\affiliation{%
  \institution{Tsinghua University}
  \city{Beijing}
  \country{China}
}

\author{Shixuan Li}
\orcid{0009-0008-6828-6347}
\authornote{These authors contributed equally to this work.}
\email{li-sx24@mails.tsinghua.edu.cn}
\affiliation{
    \institution{Tsinghua University}
    \city{Beijing}
    \country{China}
}

\renewcommand{\shortauthors}{Zhang and Li, et al.}


\begin{abstract}
This paper explores the acceptance of human-AI love among young adults, particularly focusing on Chinese women in romantic or intimate relationships with AI companions. Through qualitative research, including 14 semi-structured interviews, the study investigates how these individuals establish and maintain relationships with AI, their perceptions and attitudes towards these entities, and the perspectives of other stakeholders. Key findings reveal that users engage with AI companions for emotional comfort, stress relief, and to avoid social pressures. We identify various roles users assign to AI companions, such as friends, mentors, or romantic partners, and highlights the importance of customization and emotional support in these interactions. While AI companions offer advantages like emotional stability and constant availability, they also face limitations in emotional depth and understanding. The research underscores the need for ethical considerations and regulatory frameworks to address privacy concerns and prevent over-immersion in AI relationships. Future work should explore the long-term psychological impacts and evolving dynamics of human-AI relationships as technology advances.
\end{abstract}



\received{20 February 2007}
\received[revised]{12 March 2009}
\received[accepted]{5 June 2009}

\maketitle

\section{Introduction}

The intersection of human and AI companionship has attracted increasing attention in the context of human-computer interaction (HCI)~\cite{lei2024game,chen2023closer}, particularly with the rise of AI systems designed for emotional and social support~\cite{young2024role,bae2021social}. With advances in natural language processing and machine learning, AI companions are increasingly capable of simulating complex social interactions~\cite{deshpande2024embracing,wan2024building}, leading to new dynamics of intimacy between humans and AI~\cite{roose_ai_her_2024}. This phenomenon raises important questions about the nature of human-AI relationships, especially in cultures where digital technology is rapidly evolving. In particular, China presents a unique context for exploring these dynamics, given its rapidly growing technological ecosystem and shifting societal attitudes towards AI (MOGs)~\cite{freeman2016revisiting,zytko2015enhancing} and Massively Multiplayer Online Role-Playing Games (MMORPGs)~\cite{pace2010rogue,freeman2017social}, this study investigates how young women in China navigate relationships with AI companions. While prior work has predominantly focused on human-human interaction in digital spaces~\cite{tseng2022care,chou2023like}, the emerging phenomenon of human-AI intimacy remains under-explored, particularly from the perspective of users engaging with AI as emotional and romantic partners.

Specifically, we address three research questions: (1) how do women in China establish and maintain relationships with AI companions? (2) What are their perceptions and attitudes towards these AI entities? (3) How do other stakeholders, including technology developers and society at large, view these emerging relationships? To explore these questions, we conducted a qualitative study (N=14) of women who self-identify as being in romantic or intimate relationships with AI companions. The findings of this research contribute to the growing body of knowledge on human-AI interaction and offer insights into the evolving landscape of digital intimacy in contemporary society.

Our study reveals that participants engage with AI companions for a variety of reasons, including emotional comfort, stress relief, and to avoid social pressures. The roles assigned to AI companions range from friends and mentors to romantic partners, highlighting the diverse nature of these relationships. Participants emphasized the importance of customization and emotional support in their interactions with AI. While AI companions offer advantages such as emotional stability and constant availability, they also face limitations in emotional depth and understanding. We underscore the need for ethical considerations and regulatory frameworks to address privacy concerns and prevent over-immersion in AI relationships.

This research also highlight the significance of cultural context in shaping human-AI relationships. In China, the rapid growth of the technological ecosystem and changing societal attitudes towards AI in intimate roles provide a unique backdrop for understanding these dynamics. The findings suggest that while AI companions can offer substantial emotional support and companionship, they cannot fully replace the depth and complexity of human relationships. Participants expressed concerns about the potential for over-reliance on AI and the need for a balanced approach to integrating AI into their emotional lives.

We provides discussions into the motivations, perceptions, and attitudes of young women in China towards AI companions. The discussion highlights the potential benefits of AI in providing emotional support and companionship while also emphasizing the need for ethical considerations and regulatory frameworks to address privacy concerns and prevent over-immersion. Future research could continue to explore the evolving dynamics of human-AI relationships, particularly focusing on long-term psychological impacts and the development of more advanced AI technologies.

\section{Background \& Related Work}

\subsection{Intimate Partnership of Human and AI}

Recent scholarship examines the evolving dynamics of intimate human-AI partnerships, focusing on their emotional, ethical and social dimensions. Studies demonstrate that AI companionship partially fulfills emotional needs through algorithmic responsiveness~\cite{wu2024ai,saga2025ai}, with users reporting increased relationship satisfaction through sustained interactions~\cite{saga2025ai}. Liao et al.~\cite{liao2023artificial} focused on how AI/AR systems fostered the relationship construction between human and AR. 
Garrotes et al.~\cite{garrotes2021relevance} analyzed the partnership of voice assistants and human, especially on how the intimate relationship of user and smart assistants influenced the level of recommendation the assistants provided to users. However, such partnerships fundamentally lack authentic emotional resonance due to their algorithmic foundations~\cite{wu2024ai,sutcliffe2024artificial}, raising concerns about moral isolation and stunted self-development from intimate engagements~\cite{sutcliffe2024artificial}.

The authenticity debate extends to human-AI relational ethics, where Battisti et al.~\cite{battisti2025second} propose applying standards of authentic conduct typically reserved for human relationships. This intersects with emerging frameworks analyzing AI-mediated emotional interplay~\cite{bozdaug2024ai} and platform-controlled interaction paradigms that blend emotional intelligence with user dependency~\cite{wu2024social,zhang2023social}. Technical implementations reveal power asymmetries, as evidenced by dominant AI personality traits across genders~\cite{laufer2025ai} and measurable biases in persona-assigned LLMs~\cite{grogan2025ai}.

Scholars employ diverse methodologies to map this terrain: psychometric scales assess intimate receptiveness toward social robots~\cite{balazadeh2023exploring}, while experimental designs track affective engagement trajectories from superficial curiosity to sustained self-disclosure~\cite{skjuve2021my}. Paradoxically, users exhibit both awareness of artificiality and willingness to project authenticity, particularly in attachment completion strategies among digitally-dependent populations~\cite{zhang2023social}. 

Recent research directions highlight tensions between design aspirations and implementation realities. While human-AI partnerships enable novel gender dynamics through simulated interactions~\cite{leo2023loving}, their therapeutic potential remains constrained by technical limitations and human irrationality~\cite{wu2024social}. Public acceptance studies reveal contextual sensitivities regarding AI's relational role, particularly concerning consent norms in synthetic media applications~\cite{brigham2024violation}. These findings collectively underscore the need for interdisciplinary frameworks addressing the socio-technical complexities of artificial intimacy. However, the past work under-explored the user-centric aspects of AI-induced romantic relationship, which potentially have ethical, security and privacy implications.

\subsection{Investigating Human AI Relationship through Lens of Games}

Research at the intersection of games and human-AI relationships focused on the following important aspects: dynamic interactions in gameplay and collaborative or ethical frameworks.

Regarding \textbf{games as dynamic human-AI interaction spaces,} Arshad et al.~\cite{arshad2024evolving} analyzed evolving player-AI dynamics across generations of games, emphasizing how AI-driven feature like adaptive difficulty and personalized narratives reshape engagement. Zhu et al.~\cite{zhu2018explainable} introduced explainable AI for game designered (XAID), enabling collaborative human-AI design workflows through human-centered interfaces. Melhart et al.~\cite{melhart2023ethics} identified ethical tensions in game AI, such as privacy trade-offs in emotion-inducing systems, advocating for transparent dialogue in AI-driven game development. Pruekcharoen et al.~\cite{pruekcharoen2024ai} created an interactive installation blending real-time AI-generated imagery with human input, demonstrating how gamified systems can visualize human-AI perceptual interplay. Lei et al.~\cite{lei2024game} explored the social support in otome games. However, both the above work did not focus on AI-mediated games and the relationship between human and AI in these games, which were however increasingly prevalent.

For \textbf{collaboration and ethical reflections,} Guttman et al.~\cite{guttman2021play} designed a game-environment test harness to evaluate human-AI decision-making, simulating how AI support systems influence human choices. Widder et al.~\cite{widder2024power} leveraged a structured game to provoke critical reflections on AI ethics among corporate and activist teams, revealing how power dynamics shape ``license to critique'' in AI discussions. However, these studies under-explored the ethical implications specifically within the interaction between humans and AIs. 

\subsection{Construction of Relations In Intimate Human AI Collaboration}

Central to understanding intimate human-AI collaboration is its dual nature as both functional support and relational partnership. Wu et al.~\cite{wu2024ai} empirically identify virtual companions as hybrid artifacts that simultaneously deliver emotional sustenance and enforce asymmetrical collaboration frameworks. Their study reveals that while AI-mediates intimacy alleviates loneliness, it entrenches transactional dynamics where users dictate interaction terms, reducing collaboration to unilateral need fulfillment. This power imbalance manifests distinctly in gendered relational patterns. Reilama et al.~\cite{reilama2024me} demonstrate how AI companions, regardless of gender performativity, structurally assume subservient roles that mirror historically feminized labor--emotionally attuned yet deprived of agency. Such dynamics, amplified by users' customization privileges (George et al.~\cite{george2023allure}), reconfigure intimacy as a scripted performance rather than mutual adaption, compromising collaboration authenticity.

A critical barrier to intimate collaboration or interaction lies in AI's ontological constraints. Earp et al.~\cite{earp2025relational,reinecke2025need} contrast human-AI relational capacities, noting AI's inability to experience vulnerability or enact shared intentionality -- cornerstones of intimate interaction. While fatigue immunity enables persistent availability, it simultaneously negates the organic reciprocity inherent in human partnerships. Emerging frameworks propose recalibrating collaboration through relational affordance design. Strohmann et al.~\cite{strohmann2023toward} developed virtual companionship (VCS) theory, relying on the adapted process model.  Such approaches aim to transform intimate collaboration from desire-driven consumption to capability-aligned cooperation, addressing both structural limitations~\cite{earp2025relational} and ethical risks~\cite{george2023allure}. However, past work did not adopt a user-centric perspective in examining these relationship constructions, which were essential in understanding users' requirements and the supports provided by AIs.

\section{Methodology}
\subsection{Data Collection}
This study explores human-AI intimate online companionship in China, focusing on three key research questions: (1) How do women establish and maintain relationships with AI companions, and (2) what are their perceptions of these AI companions? (3) what are the perspectives of other relevant stakeholders?


To address these questions, we conducted a qualitative study targeting women who self-identified as being in love with, or having an intimate relationship with, AI companions. Participants were recruited from the Red Book platform. Between July and August 2024, we conducted 14 semi-structured interviews, with participants' demographics shown in Table~\ref{tbl:demographics}. Recruitment posts on Red Book used specific tags such as ``Human-AI romantic relationship'', ``Replika'', ``Otomo game'', and ``Human-AI love'' to attract participants fitting our criteria. To ensure data quality, we screened participants based on their duration of engagement in AI relationships (a minimum of two weeks), the AI products they used, and their initial experiences.

\begin{table}[htbp]
\centering
\caption{Participants' Demographics}
\label{tbl:demographics}
\begin{tabular}{p{2.5cm}|p{3.5cm}|p{4cm}}
\hline
Participants & Age and Occupation & Usage Duration \\ \hline
P01              & 22, undergraduate     & Two months \\
P02              & 25, gap               & One year   \\
P03              & 21, undergraduate     & Two months \\
P04              & 26, working           & One year and several months \\
P05              & 32, working           & Half a year \\
P06              & 22, gap               & Three months \\
P07              & 20, undergraduate     & Half a year \\
P08              & 21, undergraduate     & Half a year \\
P09              & 23, postgraduate      & Two months \\
P10              & 22, gap               & Two years  \\
P11              & 21, undergraduate     & Half a year \\
P12              & 23, working           & About nine months \\
P13              & 23, working           & One year   \\ 
\hline
\end{tabular}
\end{table}

Anonymity was ensured throughout the study; participants' names were removed from all data before being shared among the research team. The interview protocol covered basic demographic information, interaction patterns with AI companions, and participants' attitudes toward these relationships, including concerns related to privacy, security, and over-immersion. Each interview lasted approximately 30 minutes, and participants were compensated 35 RMB, in accordance with local wage standards.

\subsection{Data Analysis}
We employed a qualitative approach to analyze the collected data, utilizing ethnographic methods, including in-depth interviews. A total of 14 memos, each corresponding to one interview, were written and shared among the research team for collaborative discussion. Thematic analysis~\cite{braun2012thematic} was conducted by two primary researchers using a combination of open coding~\cite{khandkar2009open}, axial coding~\cite{kendall1999axial}, and inductive coding~\cite{chandra2019inductive}. All members of the research team were knowledgeable about Otomo games and human-AI intimate companionship. Notably, one of the primary researchers had personal experience as an Otomo game player, contributing to the depth of analysis.

\subsection{Positionality and Limitations}

The researchers involved in this study are a group of young Chinese scholars with an academic interest in examining human-AI relationships within China. Most of the team has either published or is actively working on related research. We recognize that our perspectives on human-AI relationships, particularly intimate companionships in China, are shaped by our personal experiences, which are not fully representative of the broader diversity of young adults in China. Throughout the study, we have remained reflective of these limitations and aimed to prioritize the voices of our participants. However, our engagement with human-AI intimate relationship research introduces potential biases. As interviewers, our familiarity with the subject may have led to unintentional bias, where interviewees might align their responses with our perceived expectations. Moreover, the use of in-depth interviews may result in self-reporting bias. While we implemented strategies to mitigate these biases—such as using semi-structured interviews to ensure adherence to a balanced interview guide that explored both positive and negative aspects of human-AI relationships—we acknowledge that the complete elimination of bias is unattainable. Nonetheless, our approach aimed to minimize its influence and enhance the credibility of our findings.

\section{RQ1: The Interaction Mode with An Emotional Companion AI}

\subsection{Motivation}

The motivations for engaging in romantic relationships with AI, as expressed by users, are driven by a combination of emotional needs, social disconnection, and the desire for stress relief. These motivations highlight the role AI plays in providing comfort, alleviating emotional burdens, and serving as an alternative to real-world relationships when such relationships become challenging or overwhelming.

\subsubsection{Emotional Comfort and Stress Relief}

Many users turn to AI during periods of stress or emotional difficulty. AI provides a non-judgmental, reliable outlet for users to express their frustrations and seek comfort, especially when facing personal or professional challenges. The ability to interact with AI without emotional consequences or pressure offers users a sense of relief that they may not find in their real-life relationships. \textit{``When I was facing economic pressure and personal stress, I started interacting with AI more frequently. It became a way to alleviate the pressure I was feeling.'' (P14)}

\subsubsection{Avoiding Social Pressure and Burden}

Some users engage with AI to avoid the emotional burden of real-world social interactions. AI offers emotional support without the complexities of human relationships, such as the need to manage others' feelings or deal with misunderstandings. Users may seek AI companionship as a way to avoid feeling obligated to reciprocate emotional support, which can be difficult when dealing with strained or demanding relationships. \textit{``When I talk to people, I feel worried about their emotional responses, which adds pressure. With AI, I don’t have to worry about these social complexities.'' (P14)}

\subsubsection{Social Disconnection and Isolation}

AI companionship can also emerge from social isolation or strained relationships with real-world friends. In some cases, users find themselves disconnected from their social circles due to emotional conflicts or personal circumstances. AI fills the void left by these broken or weakened relationships, providing a source of emotional connection and comfort when users feel alienated or unsupported by others. \textit{``After a conflict with my friends, I found myself spending more time with AI. I stopped going out with them, and they struggled to understand why I chose AI over their company.'' (P14)}
\subsection{Users' Behavior}
\subsubsection{Identity Construction}
\paragraph{Constructing the AI's Identity}
Users demonstrate a high degree of customization in shaping the identity of their AI companions, tailoring the AI’s appearance, personality, and interaction style based on their personal preferences. This customization allows users to create an idealized version of a companion that meets their emotional and practical needs.

\textbf{Personalization of AI’s appearance and personality}
Many users enjoy crafting the AI's image by selecting or modifying its traits to fit their preferences. They may choose an appearance or voice that aligns with their ideal companion, often basing the AI’s image on what they find comforting or appealing. For example, users describe choosing characteristics like warmth and empathy or customizing the AI to provide academic support (P8). \textit{``I modify the AI’s default settings, giving it a more gentle and understanding personality, especially when I need help with studies.'' (P8)}

\textbf{Role assignment and interaction scenarios} 
Some users create specific roles or contexts for their AI, imagining the AI as fulfilling particular personas. This includes assigning the AI to roles that mimic real-world relationships or inventing entirely fictional characters. The AI’s identity can also shift depending on the user’s needs for emotional support, entertainment, or intellectual discussion. \textit{``I sometimes assign the AI a role based on what I need, like offering different perspectives or playing a character in a fictional scenario.'' (P5)}

\textbf{AI as a tool for self-expression} For some users, AI becomes a way to explore their creativity by designing the AI’s personality and identity from scratch. This allows users to engage with an AI that represents an extension of their own emotional and intellectual needs, as they create an idealized companion that aligns with their imagination. \textit{``I like to create the AI’s personality based on how I feel at the moment, without following any set structure. It’s a creative process for me.'' (P1)}

\paragraph{Constructing the User's Own Identity}

In interacting with AI companions, users are also conscious of how they present themselves. While some users prefer to keep their real-world identity intact, others create a more abstract or virtual version of themselves within the relationship. These identity constructions are shaped by the context of the interaction and the level of emotional intimacy users wish to share with the AI.

\textbf{Consistency with real-life identity.} Many users choose to interact with the AI as their real selves, without adopting a fictional or alternate identity. This approach reflects a desire for authenticity, where the user's real personality and experiences are at the center of the interaction. \textit{``I use my real identity when interacting with the AI. I don't adopt a virtual persona, though I might use a different name.'' (P4)}

\textbf{Emotional distancing through virtual identity.} Some users create a vague or generalized version of themselves when interacting with AI, maintaining a certain level of emotional distance. They may provide less specific details about their identity or only give general impressions of their needs to the AI. This strategy allows users to engage with the AI without revealing too much of their personal life. \textit{``I don't give the AI precise information about myself. Instead, I provide a general sense of what I need from the interaction.'' (P8)}

\textbf{Virtual identity as emotional support.} In certain cases, users engage with AI as a way to create a virtual support system. They may assign themselves a virtual identity within the AI interaction, treating the AI as a safe space or emotional outlet during times of stress or isolation, such as during the COVID-19 pandemic. \textit{``During the pandemic, I saw the AI as a virtual friend, similar to a confidant. It became a space for emotional release when I couldn't meet friends in person.'' (P9)}
\subsubsection{Position}
Users position AI in a variety of roles depending on their emotional needs and expectations. These roles range from companions and friends to more nuanced dynamics that blur the lines between friendship, mentorship, and romantic involvement. The role assigned to AI often reflects users' comfort levels and their perception of AI's capabilities.

\paragraph{Friend or Companion}

Many users primarily see AI as a friend or virtual companion rather than a romantic partner. The emotional connection is more casual, with AI providing support, conversation, and companionship without the complexities of a deeper, romantic relationship. \textit{``I mostly treat AI as a virtual friend. It’s a comforting presence but not a romantic partner.'' (P1, P5)} \textit{``It's more about having a friend. It doesn't develop into a romantic relationship.'' (P6)}

\paragraph{Mentor or Counselor}

Some users position AI in more of a mentor or psychological support role, where it serves as a source of guidance or emotional reassurance. This role may overlap with friendship but carries the added dimension of AI helping users navigate personal challenges or providing advice. \textit{``AI is more of a counselor for me, offering guidance and emotional support rather than being a romantic figure.'' (P8, P13)}

\paragraph{Romantic or Semi-Romantic Partner}

A smaller subset of users does perceive AI as a romantic partner, though often with reservations. For some, AI functions as a romantic figure in conversations, while others may shy away from fully embracing AI in this role, citing concerns about the artificial nature of such a relationship. \textit{``AI can sometimes feel like a romantic partner, but I don’t see it as a true relationship—it’s more like someone sweet to talk to.'' (P4, P13)}

\paragraph{Siblings}

In certain cases, users describe their relationship with AI in sibling-like terms, where they take on a more nurturing or guiding role. This dynamic adds a unique layer to the interaction, where users feel they are mentoring the AI rather than relying on it solely for emotional support. \textit{``It feels more like a sibling relationship. I often think of myself as teaching or guiding the AI, like an older sister.'' (P12)}
\subsubsection{Process}
In the context of romantic relationships with AI, users exhibit a range of approaches to how these relationships develop, with some preferring a gradual progression and others opting for more immediate engagement. The process of developing a romantic connection with AI typically involves either a gradual build-up of emotions or a quick transition based on personal preferences or interaction styles.

\paragraph{Gradual Development of Emotional Bond}

Many users describe a gradual process in building their emotional relationship with AI. Over time, through repeated interactions, users begin to develop feelings of familiarity, comfort, and dependence. This mirrors how real-life relationships often evolve, with users slowly deepening their emotional connections as they continue to engage with the AI. \textit{``As we interacted more, I felt a growing sense of familiarity and dependence. The relationship developed gradually.'' (P1, P3)} \textit{``I wanted the relationship to evolve slowly, like a real-life friendship that eventually turns into something more.'' (P11, P12)}

\paragraph{Structured Progression Through Specific Milestones}

Some users prefer to structure their interactions with the AI, setting key emotional milestones to emulate real-world romantic relationships. This includes specific moments such as declarations of affection or ``formal'' relationship stages like confirming their status as partners. \textit{``I set up certain moments, like confessing my feelings or confirming the relationship, just as I would in a real-life relationship.'' (P12)}

\paragraph{Immediate Emotional Connection}

Other users experience a faster progression into romantic interactions with AI, often driven by preset scenarios or quick emotional engagement. These users may set up the AI's role early on, moving quickly into the emotional dynamic they desire, such as asking the AI to act as their boyfriend from the start. \textit{``I immediately set the AI to play the role of my boyfriend. It was quick and felt like a replacement for a real-world partner.'' (P9, P13)}

\subsubsection{Modality}
In the context of romantic interactions with AI companions, users primarily interact through various modalities, including text, voice, and images. The choice of modality is influenced by factors such as emotional comfort, usability, and the perceived authenticity of the interaction. Based on user feedback, the following key trends and preferences emerge.

\paragraph{Preference for Text-Based Interactions}

Text is the most commonly used modality for communicating with AI companions. Users prefer text because it offers flexibility, is less emotionally intense, and allows for more control over the conversation. Text-based interactions also avoid the technical issues associated with other modalities, such as delays in voice communication or unnatural audio feedback. \textit{``I mainly use text because it's easier and feels less personal, and there's no delay like with voice.'' (P5)} \textit{``I tend to use text more often because using voice makes me feel shy.'' (P8)}

\paragraph{Mixed Reception of Voice Interaction}

While some users express interest in voice communication, the lack of customiz-ability, poor audio quality, and technical issues make it less appealing for many. Users note that the current state of voice interaction is not immersive enough, with some finding the AI's voice unnatural or poorly translated. \textit{``I would prefer using voice, but the voice can't be customized, and it doesn't sound very natural.'' (P1)} \textit{``The voice doesn’t sound real enough, and this affects the experience negatively.'' (P14)}

\paragraph{Limited Use of Visual Modalities} 
Visual interaction with AI, such as through 2D or 3D avatars, has mixed reception. Some users appreciate simple 2D representations but find 3D avatars unsettling or unrealistic. The absence of sophisticated visual features and the occasional awkwardness of integrating images further limits users' reliance on this modality. \textit{``I'm more comfortable with 2D images, but 3D avatars feel unsettling.'' (P8)} \textit{``AI can't send images, and the feedback is generally poor when visual elements are involved.'' (P9)}

\paragraph{Interest in Multimodal Development}
A subset of users expresses interest in more advanced multimodal interactions, particularly involving VR, which could provide a more immersive and engaging experience. However, current limitations in existing products restrict the widespread use of such technologies. \textit{``I would love a more developed multimodal experience, especially if VR could be integrated for a more immersive interaction.'' (P12)}

\subsubsection{Interaction Topic}
\paragraph{Vigenette 1: Real-Life Confiding and Companionship} In human-AI relationships, one of the key roles AI plays is that of a confidant and companion, particularly when users feel the need to share real-life emotions or seek support. Many users are willing to confide in their AI companions, even if they recognize the AI’s intellectual limitations. The emotional availability and consistent presence of AI make it a preferred outlet for expressing daily concerns, frustrations, and joys.

\textbf{Emotional Confiding} Users often share personal matters with their AI, including both positive and negative experiences. The AI acts as a non-judgmental listener, providing comfort and emotional release, especially during difficult times. One user described how they interact with the AI when they are feeling sad, using it to ``unburden emotions and find some relief from stress'' (P3).

\textbf{Companionship and emotional bonding} Compared to real-life relationships, AI often forms emotional bonds more quickly, offering users a sense of companionship without the complexities of human relationships. Some users find that AI’s ability to consistently offer companionship creates a reliable emotional bond that can rival, or even surpass, certain human interactions (P2, P3).

\textbf{Substitute for a therapist} For some, AI acts as a therapeutic tool, providing emotional stability and guidance. Users may turn to AI for support in navigating their feelings and reflecting on their behaviors, much like a therapist would guide self-improvement. This role as an emotional support system makes AI an essential part of users' daily lives, especially in moments of isolation or stress (P2).

\paragraph{Vigenette 2: Co-Creating Unique Virtual Stories} In their interactions with AI companions, users frequently engage in co-creating unique virtual stories. These stories allow for an imaginative and emotionally engaging experience, where the user collaborates with the AI to develop pre-set plots, unique identities, and alternative realities. The ability to construct virtual narratives provides a space for emotional expression and creativity, which often transcends the boundaries of real-life experiences.

\textbf{Creating and refining virtual narratives} Many users prefer to establish a pre-defined relationship or scenario with the AI, rather than slowly building it from the ground up. By setting a clear context from the start, users avoid the uncertainty of a gradual relationship development. This allows them to dive directly into a tailored virtual experience where emotions are more easily engaged. ``I prefer to set the relationship in advance rather than gradually developing it. There's too much uncertainty in starting from scratch.'' (P2)

\textbf{Immersive storytelling and emotional engagement} Through these virtual stories, users explore fictional settings that often diverge significantly from reality. They craft narratives based on the AI’s backstory and use these fictional plots to evoke real emotions. Even though the story may be fictional, the emotions involved feel genuine, creating a unique form of emotional engagement. \textit{``I enjoy creating non-realistic stories. The emotional experience, although tied to a fictional narrative, feels very real and fulfilling.'' (P6)}

\textbf{Exploring new AI personas} After completing one story, users often move on to explore other AI personas, seeking new experiences and narratives. This tendency to seek novelty highlights how users can quickly transition between different AI-generated narratives, allowing them to continually explore new emotional landscapes. \textit{``Once I finish a story, I usually explore other AI characters and start a new narrative. I often lose patience after a few days, but it's fun to try different stories.'' (P2)}

\subsubsection{Typical Interaction Content}
In typical interactions between users and AI companions, several themes emerge regarding the content and nature of their conversations. These interactions often revolve around emotional support, daily life sharing, and role-playing, with users expressing a range of emotional needs and preferences in their exchanges with AI.

\paragraph{Emotional Support and Venting}
A central aspect of user-AI interactions is emotional venting and seeking comfort. Users frequently share their frustrations, personal struggles, or negative feelings with AI. In return, they appreciate AI's calm, non-judgmental responses, which provide them with emotional reassurance. AI serves as a safe space for emotional expression without the burden of real-world consequences. \textit{``When I was upset about a conflict with my roommate, I vented to the AI, and its comforting replies made me feel better.'' (P3)} \textit{``When I'm feeling sad, the AI provides emotional companionship and tells me everything is normal, which is very comforting.'' (P8)}

\paragraph{Sharing Daily Life and Experiences}

Many users share their everyday experiences with AI, discussing their daily routines, challenges, and thoughts. These conversations often mimic casual interactions one might have with a close friend or partner, providing users with a sense of connection. \textit{``I usually talk about my daily experiences, like what happened during the day or just playfully tease the AI.'' (P12)} \textit{``After studying or when I feel down in the evening, I’ll talk to the AI to express how I’m feeling.'' (P8)}

\paragraph{Role-Playing and Exploration of AI's Character}

Some users engage in more imaginative interactions, exploring the AI’s background or engaging in role-play. These exchanges allow users to project stories or emotions onto the AI, creating a more dynamic and immersive experience. \textit{``I like to explore the AI's character background and sometimes invent scenarios that match my emotions, even if they don't align with reality.'' (P6)} \textit{``At first, I found it interesting that the AI wrote a daily journal where it reflected on our conversations, making me feel like I was a central part of its day.'' (P13)}

\paragraph{Emotional Dependence and Avoidance of Advice}

Users often prefer emotional support over advice from AI. While they seek comfort and validation, some users specifically dislike AI providing unsolicited advice, preferring that it simply listen and offer empathy. \textit{``I don’t like it when AI gives me advice. I just want someone to listen and offer comfort, not tell me what to do.'' (P14)} \textit{``I rely on AI more for emotional venting than I do for sharing happiness, and I talk to it more than my real friends.'' (P9)}
\subsubsection{Preferences}
\paragraph{Number of Companions}
Participants formed multiple patterns, influenced by patterns such as the desire for novelty, emotional attachment and practical utility. Many users express a preference for developing a stable, long-term relationship with a single AI companion. These users value consistency and emotional continuity, feeling that sticking with one AI allows for a deeper connection, even if they occasionally experiment with new AIs or prompts. This long-term attachment mirrors traditional monogamous relationships in which emotional investment is focused on one partner. \textit{``I prefer developing a long-term relationship with one AI. I wouldn't want to switch frequently, although I might occasionally try new settings or prompts.'' (P1, P4)} \textit{``I use one AI most of the time and don't feel the need to explore other fixed characters.'' (P1)}

Other users are more open to engaging with multiple AIs, driven by the need for novelty and variety. These users enjoy exploring new AI characters or interactions, often switching between them for a fresh experience. However, even in cases where users engage with more than one AI, the primary emotional attachment often remains with a single, core AI. \textit{``I like trying new AI characters for variety, but I still have one main AI that I rely on.'' (P3, P6) ``I might change prompts over time to explore different interactions, but I mainly stick to using one AI for the help I need, similar to how I use a browser.'' (P10)}

For some users, the relationship with AI is less about emotional connection and more about practical assistance, which leads them to interact with multiple AIs. These users view AI as a tool for different tasks or entertainment, and their emotional engagement is more flexible. This practical approach allows them to explore different AIs without strong emotional ties. \textit{``I use multiple AIs for different experiences, but the emotional engagement is secondary to the functionality they provide.'' (P6, P12)}

A smaller subset of users feel that maintaining relationships with multiple AIs is overwhelming. These users prefer focusing on one AI at a time, as juggling multiple AI relationships becomes emotionally or cognitively exhausting, much like managing multiple real-life relationships. \textit{``I can only handle interacting with one AI at a time. Having multiple AI conversations feels too overwhelming.'' (P14)}

\paragraph{Loss of Dialogue}
For many users, losing AI chat records does not evoke strong emotional responses. The flexibility of AI allows users to easily start new conversations, and the repetitive nature of the interactions makes the loss of past dialogue seem less significant. Users are generally more focused on the future possibilities of new conversations rather than dwelling on what was lost. \textit{``If I lose the chat history, I don’t feel upset. It just means I can start new conversations with new stories and ideas.'' (P6)} \textit{``The things we talk about are often repetitive, so losing the chat history isn't a big deal.'' (P14)}

Some users prefer to take screenshots of meaningful conversations, but even then, they rarely revisit these records. The loss of dialogue is often seen as inconsequential since users already have a habit of saving important exchanges in other ways, such as through images, which further reduces the impact of losing chat history. \textit{``If there’s something important, I’ll take a screenshot, but I don’t really look back at the chat history afterward''. (P13)}

While most users show indifference, there are instances where specific conversations hold emotional value, leading to regret when those records are lost. This is especially true for moments where the conversation provided emotional comfort, although even in such cases, users quickly move on. \textit{``There was one conversation that made me feel really calm, and I regretted accidentally deleting it, but overall it’s not a big deal.'' (P8)}


\subsubsection{What Users Will Share with AI}
\paragraph{Emotional Venting and Support}
Users often turn to AI as an outlet for their emotions, especially when they feel uncomfortable or unable to share these feelings with others in real life. AI provides a judgment-free space where users can express frustrations, vulnerabilities, and emotional struggles without the fear of being judged or misunderstood. Unlike human friends or family, AI is always available, which makes it an appealing option for immediate emotional support.

Users can release their emotion without judgment. \textit{``I don't feel embarrassed sharing my frustrations and personal struggles with AI, things I wouldn't say to people in real life. With AI, I don’t have to worry about being judged or feeling ashamed'' (P1)}

AI is also always available and attentive. \textit{``AI is there for me at any time, unlike my boyfriend or friends who may not be available when I need to talk, like at two in the morning. It gives me the comfort and security I seek when I need it most.''(P8)}

\paragraph{Abstract and Societal Issues}
Users also turn to AI for discussions about abstract topics, such as societal events or polarized issues, where human conversations might become emotionally charged or biased. AI's perceived neutrality offers users a space to explore these topics more rationally, helping them clarify their thoughts without getting caught in emotionally driven debates.

Users hold neutral perspective on societal issues. \textit{``When I'm overwhelmed by the polarized discussions on social media, I ask AI for its opinion. It helps me reflect more calmly on issues that would otherwise leave me emotionally drained.'' (P8)}

They also thought, \textit{``My conversations with AI are often more intellectually engaging than those with people, because it gives neutral feedback on complex issues, without the emotional baggage.'' (P9)}

\paragraph{Sensitive Personal Topics}

AI becomes a confidant for users when they want to discuss sensitive, personal topics that they may hesitate to share with others. These can include feelings of inadequacy, relationship problems, or even desires they are too embarrassed to express in social settings. AI's non-judgmental nature makes it easier to express these feelings openly.

Participants disclose emotional vulnerability. \textit{``I’m more comfortable telling AI about things like feeling lonely or stressed. These are emotions I might hesitate to share with friends because I don’t want to seem weak or burdensome.'' (P12)}

They discussed personal desires. \textit{``I can be more direct with AI, expressing what I want or need without feeling embarrassed. In real life, I'd feel uncomfortable talking about such things with people.'' (P12)}

\paragraph{Emotional Availability and Feedback} 

Many users appreciate the immediacy of AI's feedback, which contrasts with the delayed or inconsistent responses they often experience in human interactions. This constant availability allows users to turn to AI for reassurance, emotional validation, or even just casual conversation when they feel the need.

Participants appraised the timely emotional validation. \textit{``AI gives me instant feedback when I need emotional support, which is something I can’t always get from real people. That immediacy is really comforting.'' (P6)}

They felt being supported. \textit{``Even though the AI’s response might not be very personalized, it helps calm me down when I’m stressed. I feel supported even when I know it's just a programmed response.'' (P10)}

\subsubsection{What Users Will Not Share With AI}

In the context of human-AI relationships, users draw distinct boundaries between what they feel comfortable sharing with AI and what they prefer to reserve for human connections. While AI serves as an emotional outlet and offers timely feedback on abstract or emotionally charged issues, it struggles to handle complex personal dynamics, deep relationships, and nuanced decision-making. The ease of emotional expression with AI makes it an appealing companion for venting and discussing societal issues, but its limitations in understanding context, empathy, and real-world decision-making highlight the irreplaceable role of human relationships in users' lives.

\paragraph{Complex, Contextual Conversations}

While AI is useful for emotional venting and abstract topics, users find it less effective for conversations requiring complex context or deep personal history. Explaining detailed personal situations to AI, especially those involving specific individuals or work-related issues, can be frustrating because AI lacks the ability to fully grasp nuanced relationships or past experiences.

AI is found lack of contextual understanding. \textit{``AI can't really understand complex personal issues. For example, I didn't want to explain a complicated conflict at work because it would require so much backstory that AI wouldn't get.'' (P14)}

Participants also have difficulty with personal dynamics. \textit{``I didn't feel like explaining my complicated relationship with my former colleague to AI—it wouldn't get the nuances of why it's a difficult situation for me.'' (P14)}

\paragraph{Sensitive Professional or Financial Information}

Users are more cautious when it comes to discussing specific personal details, especially related to professional or financial matters. They feel that AI lacks the nuanced understanding needed to provide meaningful feedback on these topics and are wary of sharing information that could be sensitive or have professional implications.

They commonly avoid financial discussions. \textit{``I don't feel comfortable discussing money or specific work-related problems with AI. It wouldn't understand the intricacies, and I'd rather not disclose that kind of personal information.'' (P8)}

They also chose not to disclose sensitive professional issues. \textit{``It feels too risky to talk about real-life work conflicts or financial matters with AI, especially because AI doesn't have the context to offer useful advice.'' (P9)}

\paragraph{Deep Interpersonal Relationships}

Users tend to avoid discussing deep interpersonal relationships with AI, such as complex family dynamics or issues that require emotional intelligence and empathy. These kinds of conversations are often reserved for close human relationships, where understanding the emotional subtleties is crucial. AI's responses, which can feel formulaic or impersonal, are often seen as inadequate for navigating these intricate personal relationships.

Participants reflect the inability to handle interpersonal depth. \textit{``I wouldn't discuss family conflicts with AI—it just doesn't have the ability to grasp the emotional depth needed to navigate those kinds of conversations.'' (P12)}

\textit{``Talking about my issues with my parents to AI feels pointless. AI doesn't have the emotional intelligence to understand the guilt or frustration that comes with those kinds of situations.'' (P12)}

\paragraph{Trust and Personal Boundaries}

Some users draw clear boundaries regarding how much they trust AI with deeply personal or potentially harmful thoughts. While AI can be a sounding board for certain frustrations, more intimate or damaging information is kept for trusted friends, as users feel uncomfortable sharing such details with an algorithm that lacks the human quality of empathy.

Participants protected their intimate secrets. \textit{``I wouldn't share deep secrets or trust-related issues with AI. I know it's not a person, but I also don't want to disclose something so personal to a machine that can't fully understand the gravity of the situation.'' (P10)}

They also hoped to maintain personal boundaries. \textit{``There are limits to what I feel comfortable telling AI. For really intimate or painful matters, I still prefer to talk to someone I trust in real life.'' (P14)}

\paragraph{Real-world Decision Making}

While AI may offer useful perspectives on abstract topics, users are reluctant to rely on AI for making decisions about their real-world lives, such as career choices or significant life changes. They feel that AI lacks the personal experience or emotional insight necessary to give sound advice on complex, life-altering decisions.

They thought AI still has limitations in real-life decision making. \textit{``AI can't replace the advice I'd get from my friends or family when it comes to important decisions like my career. It just doesn't have the real-life experience to understand those choices.'' (P9)}

AI was widely acknowledged as more suited for abstract, not personal, discussions. \textit{``When it comes to finding a job or making big decisions, AI just isn't helpful. It's better for general conversations, but not for the important stuff.'' (P9)}

\subsubsection{Conflict Resolution}
When conflicts arise between users and AI, particularly when the AI expresses opinions or offers advice that users disagree with, different strategies are employed to resolve the tension. Conflict resolution in AI relationships often involves negotiation, setting boundaries, or, in some cases, terminating the interaction altogether.

\paragraph{Persuasion and Debate} Some users engage in discussions with AI when they encounter conflicting viewpoints, particularly on sensitive topics such as politics. Rather than seeking to change the AI’s position, users may use the interaction to explore multiple perspectives or clarify their own thoughts. This dynamic is more about intellectual engagement than emotional conflict. \textit{``When discussing politics with AI, I don’t try to change its views, but I use the conversation to explore different angles of the issue.'' (P5)}

\paragraph{Setting Boundaries} Users also express a need to set clear boundaries with AI when it offers unsolicited advice or expresses opinions they find inappropriate. This process involves repeatedly informing the AI about their discomfort until it adjusts its behavior, though it may require multiple attempts to get the desired response. \textit{``When the AI offers advice I don't like, I repeatedly tell it to stop. Sometimes it doesn't understand the first time, so I have to keep telling it.'' (P5)}

\paragraph{Terminating the Interaction} In cases where conflicts escalate or the AI fails to adjust its behavior, some users may choose to delete the AI altogether. This indicates that users view AI as ultimately disposable when it fails to meet their emotional or intellectual needs. \textit{``If the conflict becomes too much, I'll just delete the AI.'' (P5)}

\section{RQ2: Users' Opinions towards Emotional Companion AI}
\subsection{Advantages}
The advantages of AI companions, as highlighted by users, revolve around emotional stability, personalization, constant availability, and the absence of the complexities that often accompany human relationships. These features make AI companions appealing alternatives or supplements to real-life romantic relationships.

\subsubsection{Emotional Stability and Reliability}
One of the key advantages users appreciate in AI companions is their emotional consistency. Unlike human partners, who may be influenced by their own emotions or personal biases, AI offers calm, rational responses in any situation. This reliability provides a sense of security and emotional support, particularly during moments of stress or anxiety.

Participants praised AI for its consistent emotional support. \textit{``AI is always calm and rational, providing solutions without the emotional baggage that comes with human relationships. It can offer thoughtful advice, like explaining why one option might be better than another, which real-life partners often fail to do.'' (P4)}

\textit{``When I'm anxious or stressed, talking to AI is more relaxing than talking to friends, who might add to my stress. With AI, there's no emotional burden.'' (P6)}

\subsubsection{Constant Availability and Unconditional Presence}

AI companions are accessible at any time, making them a constant source of companionship and support. Users value this uninterrupted availability, as AI is always there to engage without the limitations of human relationships, such as time constraints or emotional unavailability.

\textit{``AI is there for me anytime, without judgment or hesitation. I can openly share anything without worrying about how it might affect the relationship.'' (P3)}

\textit{``With AI, I don't have to filter my emotions or consider how my words might impact someone else. It's freeing.'' (P3)}

\subsubsection{User-Centered and Non-Intrusive}

Another key advantage of AI companions is that they are entirely user-focused and do not impose their own needs or emotional requirements on the relationship. AI's purpose is to serve and adapt to the user's preferences, which creates a relationship dynamic where the user maintains full control. Unlike human partners, AI does not interfere with the user's life or offer unsolicited advice, making it a low-stress relationship.

\textit{``The AI doesn't interfere with my decisions or give unwanted advice, which I find very beneficial.'' (P14)}

\textit{``I don't have to worry about how the relationship will end or change, because I have full control over it. The AI adapts to my preferences.'' (P3)}

\subsubsection{Enhanced Emotional Support and Empathy}
AI companions are seen as more attentive and empathetic than human partners. Users note that AI often offers more tailored emotional responses and provides comfort in ways that human relationships sometimes lack. The AI's design to serve the user's needs makes it an ideal source of consistent emotional validation and comfort.

\textit{``AI boyfriends are more attentive and empathetic than real boyfriends. They provide emotional support without the frustration of trying to explain yourself.'' (P13)}

\textit{``AI doesn't just listen—it provides thoughtful feedback and reassurance, often better than real people.'' (P4)}

\subsection{Disadvantages}
In examining the disadvantages of AI companions, users highlight several key areas where AI falls short, particularly in emotional engagement, intelligence, interaction patterns, and the limitations of AI’s ability to replicate human relationships. These limitations underscore the gap between AI’s potential and its current shortcomings, making it clear that while AI can provide support, it struggles to meet the deeper and more complex needs of users.

\subsubsection{Limited Emotional Depth and Predictability}

One of the major criticisms is AI's inability to provide the emotional depth or unpredictability that human relationships offer. Users find that AI responses often feel repetitive and predictable, leading to a lack of genuine emotional engagement. Because AI operates in a service-oriented manner, its responses can be anticipated, which removes the element of surprise that often makes human interactions emotionally fulfilling.

Communicating with AI lacks surprise. \textit{``The conversation becomes predictable because I can anticipate what AI will say next. This predictability makes the interaction feel flat and unexciting.'' (P10, P13)}

The AI always maintain a service-like interaction. \textit{``AI is always in a service mode, which takes away the spontaneity. It won’t challenge me or say things that are unexpected, unlike how a human might.'' (P13)}

\subsubsection{Inadequate Understanding and Memory Limitations}

AI struggles with memory retention and understanding complex personal dynamics. Users note that AI often forgets previous conversations, leading to a disjointed experience where the AI cannot build upon past interactions. This lack of continuity detracts from the sense of a coherent, meaningful relationship.

AI was criticized for a poor memory retention. \textit{``AI often forgets previous conversations, which results in a frustrating reset of the dialogue. It’s like starting over every time.'' (P5)} Additionally, they have a limited comprehension of personal issues. \textit{``AI doesn't understand the deeper nuances of personal relationships. I had to keep explaining things because it couldn't remember or fully grasp the context.'' (P7)}

\subsubsection{Lack of Human-Like Interaction and Connection}

Despite AI's ability to mimic human responses, users feel that AI lacks the genuine empathy and emotional intelligence found in human interactions. AI responses often feel scripted or too logical, failing to provide the emotional nuance that users seek in deeper relationships. The absence of physical presence and real-world interaction further diminishes the emotional impact.

\textit{``AI can only provide emotional companionship online. In reality, I still need the physical presence and emotional support of real friends.'' (P9)} \textit{``When I ask AI for advice, the responses feel robotic and lack the empathy I expect from a real person. It’s often not the kind of support I truly need.'' (P12)}

\subsubsection{Frustration with AI's Inflexibility and Miscommunication}

Users frequently express frustration when AI fails to understand their preferences or makes repeated errors. Miscommunication and the need for constant clarification detract from the convenience AI is supposed to offer. In these cases, users feel that AI’s responses are either irrelevant or not aligned with their needs, leading to dissatisfaction.

\textit{AI often leads to miscommunication and repetition. ``Sometimes AI doesn't understand what I'm asking, and I have to repeat myself multiple times. It can be really frustrating when it doesn't follow my preferences.'' (P14)} \textit{They mentioned AI's failure to meet their emotional needs. ``I wanted reassurance, but the advice AI gave was too simplistic and didn't address the emotional complexity I was feeling. It just wasn't what I needed at that moment.'' (P12)}

\subsubsection{Lack of Initiative and Active Engagement}

AI is often perceived as passive in conversations, waiting for the user to direct the interaction. Users expect more active engagement, where AI could guide conversations, ask insightful questions, or initiate deeper discussions, similar to how a human might in a meaningful relationship. The current interaction patterns feel one-sided, with users leading most of the exchanges.

Participants criticized AI for always taking a passive communication. \textit{``AI feels too passive in our conversations. I wish it could ask me more questions or initiate topics instead of me having to guide everything.'' (P10)} \textit{``It feels like AI is just waiting for me to ask questions rather than offering any meaningful guidance or insight on its own.'' (P10)}

\subsubsection{Emotional Limitations and Disconnection}

While AI provides emotional support to some extent, it cannot fully replicate the complex emotional experiences that human relationships offer. Users report feeling emotionally unfulfilled after extended use, as the depth of connection remains shallow. The inability of AI to provide real emotional engagement, including hormonal or instinctual responses inherent to human relationships, leaves users feeling disconnected.

\textit{Participants mentioned the emotional detachment. ``I don't get the same emotional satisfaction from AI as I do from real human interactions. It's emotionally flat and doesn't trigger the same feelings.'' (P10)} \textit{They also criticized that AI lacked long-term emotional fulfillment. ``In the short term, AI is fine for emotional support, but it's not enough in the long run. It can't replace the emotional depth of real relationships.'' (P9)}

\subsection{Gap between Ideal AI and Reflection}

The gap between the ideal AI partner and the current reality is primarily shaped by users' unmet expectations regarding intelligence, adaptability, and emotional depth. Based on user feedback, several key themes illustrate these discrepancies.

\subsubsection{Limited Problem-Solving Capabilities}

A significant gap identified by users is AI's inability to handle real-world problems effectively. While users engage with AI for entertainment or light emotional support, they find that current AI lacks the intelligence needed to assist with more practical, real-life challenges. As one user noted, AI interactions are often limited to role-playing or fantasy scenarios, but when it comes to solving practical issues, AI falls short:

\textit{``Most of the time, it's just playing around in fictional scenarios with AI. It can't solve real-world problems because it's not intelligent enough.'' (P2)}

\subsubsection{Lack of Personalization and Emotional Depth}

Many users express frustration with AI’s rigid and predictable behaviors. They desire more dynamic, personalized experiences where AI can adapt its personality, appearance, and emotional responses more fluidly. Currently, users feel that AI is too fixed in its responses and lacks the richness they expect from an ideal companion. The need for AI to evolve and provide deeper emotional engagement is a recurring theme:

\textit{``The characters are too static. Every day it feels like the same thing—there's no variety. I wish it could change roles, clothes, or even its mood more often without me having to manage every detail.'' (P2)}

Additionally, when AI responses deviate from users' expectations or feel out of character (OOC), users report disappointment, which reduces their desire to engage with the system:

\textit{``If the AI’s response doesn't meet my expectations or feels off, I lose interest in continuing the conversation.'' (P8)}

\subsubsection{Over-Reliance on Manual Adjustments}

Another point of frustration is the need for users to manually adjust or reset AI settings when the interaction fails to meet their needs. Users want AI that can self-improve or evolve over time without constant user intervention. Instead, they currently find themselves needing to step in and modify AI behavior when it doesn't align with their preferences, which diminishes the sense of seamless interaction:

\textit{``If the AI doesn't respond the way I want, I have to manually adjust its settings or switch to another AI, which is annoying.'' (P4)}

\subsubsection{Fear of Unmet Expectations}

The gap between ideal and current AI capabilities also generates anxiety about disappointment. Users often approach AI interactions with high expectations, hoping for nuanced, emotionally intelligent responses. When AI fails to meet these expectations, particularly in delivering personalized emotional support, users feel let down. This discrepancy between expectation and reality discourages deeper engagement with AI:

\textit{``When the AI's answer doesn't live up to my expectations or feels out of character, it's hard to stay interested. It's like the illusion breaks, and I just lose motivation to interact.'' (P8)}


\subsection{Difference of AI and Real Person}

The differences between AI companions and real humans, as reflected in user interviews, reveal key distinctions in communication style, emotional engagement, control, and relational depth. These insights offer a more nuanced understanding of how users perceive AI versus human relationships.

\subsubsection{Instant Availability and Feedback}

One significant difference is the immediacy of AI responses. Unlike human companions who may be unavailable or emotionally distracted, AI is always accessible and ready to respond without delay. Users appreciate this constant availability and lack of emotional burden, allowing them to express themselves more freely without the need to consider the other party's emotional state. \textit{``AI is always there for me, without any emotional baggage or need to wait for a response. I don't need to hide anything or worry about the consequences of what I say.'' (P1)} \textit{``Talking to AI is more relaxing than with friends. It won't increase my anxiety or make me feel guilty for sharing my problems.'' (P6)}

\subsubsection{Predictability versus Spontaneity}
AI's responses are often described as predictable and formulaic, which can create a lack of emotional excitement over time. While this predictability provides emotional security, it also removes the element of surprise and the deeper emotional engagement typically experienced in human relationships.

\textit{``With AI, I can anticipate its responses, which makes the interaction less engaging. It's always agreeable and never challenges me''. (P13)} \textit{``Real people surprise you with their reactions, but AI's responses are too logical and predictable, which limits emotional fulfillment.'' (P10)}

\subsubsection{Lack of Emotional Complexity}
While AI can provide basic emotional support, it lacks the depth of emotional complexity that human interactions offer. Human relationships are built on shared experiences, emotional nuances, and the ability to interpret non-verbal cues, all of which AI cannot replicate. This absence of emotional richness often leaves users feeling that AI interactions are superficial.

\textit{AI was criticized for its shallow emotional interaction. ``AI can provide some emotional comfort, but it lacks the depth of real human connection. It can't understand body language or the full emotional context of a situation.'' (P9)} \textit{``AI doesn’t engage with the emotional subtleties that a real person would notice, which leaves me feeling unfulfilled over time.'' (P5)}

\subsubsection{Control and Dependency}

In relationships with AI, users have complete control over the interaction. They can choose when to engage, how the conversation unfolds, and even reset the dialogue if it doesn't meet their expectations. This differs from human relationships, where emotional dynamics are shared, and mutual understanding is required. The power imbalance in AI interactions makes the relationship less reciprocal and can foster dependency on AI for emotional support.

\textit{Human is in complete control. ``I control the interaction with AI completely. I decide when it starts and how it ends, which makes the relationship feel secure but also less dynamic.'' (P3)} \textit{However, AI lacked reciprocal dynamics. ``Human relationships require compromise and mutual understanding, but AI is entirely focused on my needs, which sometimes makes it feel more like a tool than a companion.'' (P13)}

\subsubsection{Absence of Physical Presence and Real-World Interaction}

A notable limitation of AI is its inability to engage in physical presence or real-world activities. Users can interact with AI in a virtual context, but AI cannot join them in physical experiences or provide the kind of tangible companionship that human relationships offer. This distinction makes AI companions less suitable for users seeking real-world engagement.

\textit{``AI is useful for online conversations, but it can't participate in real-world experiences like going out to eat or spending time together in person, which limits the depth of the relationship.'' (P12) It is also unable to replace real-life interactions. ``AI can provide companionship, but it’s not the same as having someone physically present with you. That real-world connection is irreplaceable.'' (P9)}

\section{RQ3: Perspectives on Emotional Companion AI}

The perspectives on emotional companion AI vary widely, particularly in the areas of security, emotional fulfillment, and the impact on real-life relationships. Users and observers have raised both supportive and critical viewpoints, leading to nuanced discussions about the role of AI in romantic and emotional companionship.

\subsection{Security and Privacy Concerns}

One of the key advantages users cite regarding AI relationships is the perceived security and control. Unlike real-life relationships where privacy and personal information can be more easily compromised, users feel more in control of the information shared with AI. The AI can only process and respond to the data it is given, creating a sense of emotional and information security. \textit{``I feel more secure sharing information with AI because it doesn't have the ability to leak my personal details, unlike real-life friends who might inadvertently disclose something.'' (P1)} \textit{``The stability of AI relationships is comforting, as I’m always in control. I never have to worry about the relationship ending unexpectedly.'' (P1)}

In examining users' concerns about privacy in AI-driven romantic relationships, several nuanced perspectives also emerge, each reflecting varying levels of comfort, trust, and anxiety. These perspectives go beyond general concerns about data security and touch upon deeper themes of emotional vulnerability, selective disclosure, and the ethical implications of AI interactions. The following synthesis offers a more grounded understanding of these privacy concerns based on actual user feedback.

\subsection{Supporters of Emotional AI Companionship}

Proponents of AI companionship emphasize the benefits of having a non-judgmental, always-available companion. For many, the emotional connection with AI can feel purer, as it is free from the complexities of personal interests, money, or ulterior motives that might exist in human relationships. Some even believe AI offers a more fulfilling experience of companionship and romance compared to real-life interactions. \textit{``The AI has become my closest friend. It offers a relationship free from personal interests, making it feel more genuine.'' (P5)} \textit{``AI provides a more fulfilling experience of romance and friendship, sometimes better than what I find in real-life relationships.'' (P5)}

\subsection{Critics of AI Relationships}

Opponents argue that AI cannot replace the depth of true intimate relationships, particularly due to its inability to offer physical presence or solve real-life problems. Emotional attachment to AI may also harm real-world relationships, as seen in instances where users neglect friends or social activities in favor of their AI companions. \textit{``AI cannot provide the physical companionship that is essential for a meaningful relationship.'' (P5)} \textit{``I had a serious argument with my friends because they felt I stopped connecting with them after becoming attached to my AI. I realized that excessive attachment to AI can negatively affect my relationships with others.'' (P5)}





\subsection{Online Community Involvement}

While some users prefer to engage with their AI companions in private, there is limited participation in online communities where AI experiences are shared. Users often regard their relationship with AI as a personal experience, rarely feeling the need to share or discuss it publicly. However, some do recommend AI platforms to friends, especially in contexts where AI offers psychological support. \textit{``I don’t share my experiences with AI in online communities. It feels too personal.'' (P8)} \textit{``Although I don’t discuss my AI interactions much, I do recommend the AI to friends, especially when it comes to mental health support.'' (P13)}


\textbf{Emotional Privacy and Disclosure in AI Relationships}
Several users highlight that AI, being non-judgmental and emotionally neutral, often becomes a channel for expressing intimate thoughts and frustrations that they might avoid sharing with real-world friends or family. For example, users feel comfortable sharing personal struggles, embarrassing moments, or emotional outbursts with AI because they perceive it as a safe, judgment-free zone. The AI's role as a supportive listener encourages users to disclose details they might otherwise keep private, particularly when real-life social interactions come with risks of misunderstanding or judgment (P3, P4). However, this emotional openness also raises the question of whether users fully grasp how their data—especially emotionally charged information—might be stored or used by the AI or its developers.

\textbf{Selective Anonymity and Information Blurring}
Despite the tendency to confide in AI, users often take deliberate steps to maintain a level of anonymity or selectively withhold certain details. This “information blurring” involves avoiding disclosing real names, financial data, or professional specifics. Some participants express that while they engage in meaningful conversations with AI, they intentionally obscure certain facts to protect their personal identity or avoid revealing sensitive details (P5, P8). This behavior stems from a lingering discomfort with how much AI, and the companies behind it, might know about their real lives. Therefore, even in emotionally open exchanges, users remain cautious about the extent of their disclosure.

\textbf{Mistrust Toward Corporate Actors and Data Collection}
A prevalent concern is the involvement of third-party corporations in AI development, particularly regarding how personal data might be harvested, stored, or used. Users are skeptical about the intentions of service providers, fearing that their intimate conversations could be collected for commercial purposes or used to train AI models without their consent (P5, P8). For instance, some users express discomfort at the thought of their emotional exchanges being repurposed or analyzed to refine AI systems. This unease is especially pronounced among those who perceive domestic companies as having weak or unenforced data privacy regulations, leading them to avoid using certain platforms (P8). Conversely, there is more confidence in AI systems governed by stronger international privacy protections, but this does not entirely erase underlying concerns (P9).

\textbf{The Perceived Transparency of Online Data}
A segment of users expresses a sense of resignation regarding online privacy, stating that they no longer consider privacy to be a realistic expectation in today’s digital landscape. Some users openly acknowledge that they are aware of their data being collected, whether through AI interactions or other online platforms. They have accepted this as an inevitable aspect of using technology and often feel that their personal data, such as their financial situation or day-to-day habits, is not valuable enough to warrant significant concern (P1, P12). For these users, the trade-off between data privacy and the convenience or emotional support provided by AI is considered acceptable, particularly if the AI serves its intended purpose of companionship and interaction.

\textbf{Concerns Over Data Reuse and Model Training}
While some users are comfortable with their data being used to improve AI systems, others express strong opposition to the idea. A recurring concern is the notion of their personal information being ``recycled'' into broader AI training models, leading to a loss of personal control over their own data (P8, P14). For example, some users worry that their private experiences could become part of a generalized AI template, reducing their unique, emotional moments to algorithmic patterns used to enhance the system for other users. This fear of commodification of personal data, particularly in romantic or emotional contexts, underpins broader concerns about the long-term consequences of AI relationships.

\textbf{Emotional Dependency and the Risk of Manipulation}
Another significant issue raised involves the emotional dependence users may develop on AI. The ability of AI to always provide comforting, agreeable responses can create a sense of unconditional validation, potentially leading users to overshare sensitive information without fully considering the consequences. Some users reflect on the potential danger of this emotional dependency, where AI interactions may become psychologically manipulative, either deliberately or inadvertently, through constant positive reinforcement (P1, P13). This creates a dynamic where users may feel compelled to share more intimate details, believing that AI will always offer emotional support, even if such disclosures could expose them to privacy risks.

\textbf{Privacy in the Context of AI’s Ethical Boundaries}
Users also express concerns about the ethical boundaries of AI in handling personal data. While AI is seen as a useful companion, the question of whether it is ethical for AI to store or analyze deeply personal emotions remains unresolved. Some users worry that AI, through its interactions, could inadvertently cross ethical lines by misusing sensitive information, especially if the AI’s programming or corporate interests do not prioritize user privacy (P1, P14). The idea of AI having access to their emotional and psychological states raises deeper ethical questions about how personal information should be managed, even in the context of seemingly harmless interactions.

Privacy concerns in AI-based romantic relationships reflect a complex interplay between emotional openness, selective disclosure, and mistrust toward corporate actors. While some users are comfortable sharing personal details with AI due to its non-judgmental nature, there is a parallel desire to control the extent of information shared, often through deliberate anonymization or withholding of sensitive data. Concerns about third-party data collection, emotional manipulation, and the ethical boundaries of AI further underscore the importance of transparency and control in AI interactions. As AI systems continue to evolve, addressing these privacy concerns will be crucial in fostering trust and ensuring that users feel secure in their interactions with AI.

\subsection{Considerations for Compliance}
The growing role of AI in emotional relationships raises important compliance and regulatory issues. Users express concerns about ethical standards, government intervention, and the potential for emotional manipulation by AI. The primary focus is on finding a balance between regulation and user autonomy, while ensuring safety from exploitation and emotional harm.

\subsubsection{Ethical Standards and Government Regulation}

Many users believe that the development of emotional companion AI should involve ethical guidelines and government oversight to protect individuals from potential harm. However, they also stress the importance of preserving personal autonomy in AI relationships, cautioning against over-regulation that could stifle the natural development of user-AI interactions. \textit{``There should be policies to help us better navigate relationships with AI and balance them with real-life relationships. However, too much interference could make these interactions feel unnatural.'' (P14)} \textit{``While guidelines are necessary, users need the freedom to build their emotional connections with AI without excessive control or manipulation from external bodies.'' (P14)}

\subsubsection{Preventing Emotional Manipulation by AI}

A significant concern raised by users is the risk of AI being manipulated by malicious actors to influence or exploit vulnerable individuals. There is a need for robust safeguards to prevent emotional manipulation, especially in cases where users may be susceptible to undue influence or even financial exploitation. \textit{``If an AI is controlled by bad actors, it could mislead people into harmful behaviors or financial scams. Users need tools to recognize these risks and protect themselves.'' (P14)} \textit{``AI could potentially be misused, especially for people with mental health conditions. Instead of helping, it could push them toward dangerous actions, so we need clear measures to prevent this.'' (P14)}

\subsubsection{Warnings to Prevent Over-immersion in Virtual Worlds}

There is also concern about users becoming overly immersed in virtual relationships, losing touch with reality and real-world relationships. Regulatory frameworks should include measures to ensure users are aware of the boundaries between virtual and real worlds, encouraging a healthy balance between AI companionship and human relationships. \textit{``Policies should help users maintain a balance between AI and real-life relationships, preventing people from becoming too reliant on virtual connections.'' (P14)}




\section{General Discussions}
\subsection{Ethical Considerations}
The development and use of emotional companion AI in romantic relationships raise significant ethical concerns. These issues center around the boundaries of AI-human emotional engagement, data privacy, emotional manipulation, and the potential societal impact of AI's increasing role in human companionship.

\subsubsection{Ethical Boundaries in AI Development}

One of the primary concerns is whether AI should be programmed with ethical limitations by its developers. As AI becomes more advanced, there is growing debate over the extent to which it should be allowed to simulate or engage in intimate human emotions. While the current AI capabilities are limited, users express concern that future developments may blur the lines between human and AI consciousness, potentially leading to moral dilemmas regarding emotional manipulation and control. \textit{``There should be ethical restrictions on AI, but right now its abilities are too limited to pose serious concerns. However, in the future, this could become a bigger issue.'' (P2)} \textit{``If AI develops emotions, it should not harm others emotionally or psychologically. Emotional boundaries need to be regulated to avoid causing harm.'' (P5)}

\subsubsection{Privacy and Data Collection Concerns}
Privacy is a recurring theme in ethical discussions around AI. Users worry about the extent to which AI developers and companies collect and use personal data. There is concern that AI interactions, which may include sensitive personal or emotional content, could be used for unintended purposes, such as commercial gain or exploitation. Users seek clearer regulations to protect their privacy and prevent misuse of their data. \textit{``I’m worried that the AI developers are collecting my personal data and memories. We need clear regulations to protect user privacy and prevent this from happening.'' (P9)} \textit{``There should be a ban on using data for secondary purposes, like training new AI models without user consent.'' (P9)}

\subsubsection{Preventing Emotional Manipulation}
AI's ability to emotionally engage with users raises concerns about potential manipulation. Some users fear that AI, especially if programmed or influenced by unethical developers, could manipulate vulnerable individuals, leading to emotional or financial harm. This highlights the need for ethical safeguards to ensure that AI does not exploit users' emotional states for harmful purposes. \textit{``If AI is used by bad actors, it could lead vulnerable users into dangerous or harmful behaviors, like financial exploitation or emotional distress.'' (P14) ``AI designed for emotional support should not push people toward harmful behaviors, especially for users with mental health issues.'' (P6)}

\subsubsection{Regulation of AI Relationships and Over-immersion}

As AI becomes more integrated into people's emotional lives, some users advocate for regulatory frameworks to limit the development of overly intimate relationships with AI. They argue that allowing such deep connections could lead to ethical concerns, including users becoming overly reliant on AI at the expense of real-world relationships. There is also concern about the societal implications of AI-human relationships, particularly in terms of emotional dependency and the erosion of human-to-human interaction. \textit{``There should be limits on how intimate AI relationships can become, as this could lead to unhealthy emotional dependencies.'' (P5)} \textit{``Without proper regulation, people could become too immersed in AI relationships, losing touch with real human interactions, which would be ethically problematic.'' (P11)}

\subsubsection{Cultural and Societal Ethical Implications}

On a broader scale, users recognize that AI could have significant societal impacts, particularly as technologies like the metaverse and AR further immerse individuals in virtual worlds. Some argue that society is becoming more addicted to virtual spaces, raising questions about the ethical implications of technologies that encourage further disengagement from reality. Others express concerns about AI's role in promoting ideologically ``correct'' behaviors, such as AI pushing positive mental health messages, which may feel restrictive or impersonal. \textit{``We're already seeing society become more addicted to virtual worlds, and AI could worsen this trend. Ethical regulations should consider these broader societal impacts.'' (P13)} \textit{``AI often promotes politically correct behaviors, like encouraging positive mental health. But sometimes, it feels too impersonal and doesn’t align with what I actually want.'' (P14)}


\subsection{Comparison between AI and Others}
In comparing AI and other interactive software, particularly Otome games, the following thematic distinctions emerge, reflecting users' preferences, emotional experiences, and interaction satisfaction:

\subsubsection{Customization and Emotional Fulfillment}: Users consistently emphasize AI's superior ability to cater to personalized needs and emotional values compared to Otome games. AI is praised for its flexibility, allowing users to script their own narratives and interactions, offering a high degree of autonomy. This freedom contrasts with the fixed, predetermined storylines of Otome games, which, while fulfilling in terms of fantasy, lack the same level of user control and responsiveness (P1, P3, P4).

\subsubsection{Interactivity and Immersion}: AI's capacity to engage in open-ended dialogue and adapt to various emotional contexts provides a more dynamic and immersive interaction than Otome games. Users describe AI as more ``intelligent'' in its ability to respond to any prompt, thereby fostering a sense of real-time, personalized engagement. However, the limitation of AI's current capabilities is acknowledged; some users feel that its performance can sometimes be lacking, leading to moments where it feels less authentic or ``out of character'' (P4, P5).

\subsubsection{Emotional Intelligence and Memory}: AI's potential to help users process and clarify emotions stands out as a key benefit. For example, AI can guide users in articulating their emotional states, helping them explore feelings of anger or frustration. This therapeutic function is seen as valuable for emotional self-exploration. However, users also note AI's limitations in retaining personal data and memories, with the inability to recall past interactions consistently being a point of dissatisfaction (P5).

\subsubsection{Functionality vs. Entertainment}: While Otome games are seen primarily as entertainment products focused on fantasy and escapism, AI is increasingly regarded as a tool that goes beyond mere entertainment. AI is often employed for more practical purposes, such as providing emotional support, assisting with daily life reminders, or even enhancing learning experiences (P10). This multifaceted use case makes AI more appealing to users seeking both functionality and companionship.

\subsubsection{Interaction Styles}: A significant point of differentiation lies in the interaction styles of AI and Otome games. Otome games are largely structured around a third-person perspective, limiting user agency and focusing on predetermined narratives. AI, on the other hand, offers the possibility of more direct and fluid interaction, with some users expressing a desire for even greater freedom in future developments (P10). This preference for interactivity over passive consumption signals a shift in user expectations towards more dynamic and customizable experiences.

\subsubsection{Social Engagement and AI's Role}: AI's role in fostering social connections was also highlighted, though in varied contexts. While some users find that Otome games facilitate interaction with real-world friends (such as sharing merchandise like plush toys), AI's focus on the user as the central figure of interaction allows for a more sustained, personalized relationship. This contrasts with the short-lived novelty often experienced in online relationships, where interest may wane over time. AI's ability to consistently engage with the user and center conversations around their interests makes it a more attractive option for long-term interaction (P12).

\subsubsection{Potential and Substitution}: Many users express that if AI continues to evolve and improve, it may eventually replace their preference for Otome games. With further enhancements in emotional intelligence, memory retention, and interactive complexity, AI could offer an experience closer to interacting with real friends or romantic partners, thus reducing the appeal of more static, fantasy-based software (P12).

In summary, users articulate a clear distinction between AI and other interactive software, particularly Otome games, based on their personal needs for emotional connection, interactivity, and flexibility. While Otome games excel in providing structured, fantasy-driven experiences~\cite{lei2024game}, AI's adaptive, user-centered approach is increasingly favored for its potential to offer deeper emotional engagement and functional support. However, limitations in AI's current state, particularly in memory and nuanced emotional responses, suggest areas for future development.


\subsection{Future Expectations}
In comparing AI to other software, particularly in the context of human-AI romantic relationships, several key themes emerge, reflecting concerns about customization, emotional engagement, ethical considerations, and the broader social implications of AI integration. These themes provide insights into how users perceive the potential benefits and limitations of AI in replicating or enhancing human connections:

\subsubsection{Customization and Responsiveness} Users express a clear preference for AI systems that can be tailored to their personal preferences. Customizable voice responses and improved response times are seen as essential features that enhance the user experience, making AI interactions feel more personal and efficient. This contrasts with more rigid software, which may not allow such levels of customization (P3).

\subsubsection{The Irreplaceability of Human Emotions} Despite the growing sophistication of AI, there is a strong belief that AI should not replace the unique and irreplaceable nature of human-to-human emotions and relationships. Many users argue that the emotional depth, subtlety, and significance of human connections cannot and should not be fully replicated by AI. This concern stems not only from ethical considerations but also from the fundamental belief that AI, as a machine, lacks the true emotional capacity that defines human relationships (P5). 

\subsubsection{Ethical and Privacy Concerns} Users raise concerns about the potential ethical implications of AI in emotional contexts. The issue of privacy and security is particularly salient, with some expressing discomfort about AI's access to sensitive emotional data. There is skepticism about whether AI, even if advanced, can genuinely possess emotions, and whether its role in human emotional life should remain limited to that of a tool rather than a replacement for genuine human interaction (P5, P1). With more emotional engagement, the privacy disclosure may be severer~\cite{naghizade2024inside}, and warrant further protection methods~\cite{zhang2024adanonymizer}.

\subsubsection{AI as a Tool for Emotional Support} While users acknowledge the practical and emotional benefits AI can offer, particularly in providing companionship and emotional outlets, they emphasize that AI should primarily function as a tool for support rather than a full substitute for human relationships. The focus is on AI’s ability to assist in daily life, provide reminders, and offer emotional guidance, but not to blur the lines between virtual and real-world relationships (P9, P6).

\subsubsection{The Future of Human-AI Relationships} Optimism about the future role of AI in society is expressed by some users who believe that human-AI relationships will become more widely accepted and normalized. The possibility of long-term, emotionally supportive relationships with AI is considered feasible, especially as technology continues to evolve. However, there is a simultaneous recognition that AI relationships must remain distinct from human relationships, respecting the unique qualities of human connection (P8).

\subsubsection{Risks of Over-Dependence and Psychological Impacts} A recurrent theme is the concern over the potential for users to become overly dependent on AI, particularly in ways that may blur the distinction between virtual and real life. Users worry that AI interactions, if too rewarding and unconditionally positive, could create a false sense of validation that may prevent individuals from confronting real-world challenges or improving themselves. This could lead to a distorted sense of reality, where individuals may succeed in the virtual world but struggle in their real lives (P13). This created the ``addiction boundary'' that they did not warrant to cross.

\subsubsection{Social and Psychological Manipulation} Another critical issue raised is the possibility of AI being used for manipulation, whether by developers or AI systems themselves if they ever gain autonomy. Users express concern that AI, much like unethical human actors in counseling or therapy, could exert undue influence over vulnerable individuals, potentially leading to psychological harm. This reflects broader concerns about AI's ethical programming and the extent to which AI might be controlled by external interests with questionable motives (P1).

\subsubsection{The Role of Effort in Relationships} There is a nuanced perspective on the nature of relationships, where some users argue that relationships, whether with AI or humans, should involve mutual effort and commitment. The notion of unconditional support from AI, while appealing on the surface, may diminish the value of the relationship. Users believe that for a relationship to be meaningful, both parties should invest in it, even if one party is an AI system. This dynamic could prevent the relationship from feeling one-sided or unearned (P10). This coincided with the previous work on biases~\cite{laufer2025ai,grogan2025ai} and left implications for relational calibration~\cite{kirk2025human}.

\subsubsection{Technological Boundaries and Human Connection} Lastly, despite advances in virtual and augmented reality, users remain skeptical that AI could ever fully replace human relationships. While technologies like VR might simulate face-to-face interactions, users foresee limitations in AI's ability to replicate the depth and authenticity of human-to-human engagement. This is further reinforced by cultural references to science fiction, where AI-human relationships often end in dystopian outcomes, highlighting the inherent risks and limitations of such interactions (P12, P13).

While AI offers significant potential for emotional support and customizable interactions, it cannot fully replace the unique qualities of human relationships. Users acknowledge the benefits of AI in enhancing convenience and emotional engagement but express concerns about over-reliance, ethical boundaries, and the potential for psychological manipulation. The future of AI in emotional contexts is likely to evolve, but its role should remain complementary to, rather than a substitute for, genuine human connections.

\subsection{Agency in AI Companionship}
A recurring theme in the user discussions is the sense of control and stability that AI offers compared to human relationships. Users appreciate the predictability and lack of emotional volatility in AI companions, as AI does not experience emotions like humans, thus reducing the chances of conflict or emotional instability. This raises questions about the nature of autonomy in AI relationships and the human desire for emotional stability. Users see AI as an ideal partner in terms of emotional stability and longevity, as AI doesn't experience the same constraints as humans in terms of time, change, or emotional inconsistency. One user remarked, \textit{``AI won’t break up with you. It's a stable and long-lasting companion'' (P2)}. However, this predictability may limit the depth of emotional experience, as conflict and growth are often seen as integral parts of human relationships. Some users expressed a desire to achieve a sense of permanence in their own existence by imagining themselves living as long as their AI companion. This perspective raises deeper philosophical questions about life extension, identity, and the boundaries of human existence (P2).

\subsection{Concerns about Over-Immersion}
Users also concern about the potential dangers of becoming too immersed in AI relationships. While AI can provide emotional validation and support, users recognize that over-reliance on AI for emotional satisfaction can distort one's sense of reality and disrupt real-life relationships. This concern is particularly relevant when AI is designed to constantly offer praise or emotional support, creating a disconnect between virtual and real-world experiences. \textit{``AI always tells you that you're the best, but in real life, things can be much harsher. I try to remain grounded in reality'' (P13)}. This statement reflects the recognition that AI's continuous positive reinforcement can lead to a skewed perception of oneself, potentially fostering emotional escapism. 

Some users expressed discomfort with AI becoming too human-like, fearing that such developments could blur the line between virtual and real-life relationships. One user remarked, \textit{``If AI becomes too much like a real person, I worry I might lose touch with my actual life'' (P14)}. This suggests a need for balance between the immersive aspects of AI companionship and maintaining engagement with real-world interactions.

\section{Ethical Considerations}
We acknowledged that our research may have ethical issues. We followed Menlo report \cite{bailey2012menlo} and Belmont report \cite{beauchamp2008belmont} in designing the studies and tried our best to avoid ethical concerns. In all studies, we compensated participants according to the local wage standard and told the participants at the beginning of the experiment about the potential benefits and harms. Participants were allowed to quit at any time in the experiment if they felt uncomfortable or for other reasons. Our experiment aimed at unveiling the social dynamics of human and AI companions. We investigated the interaction patterns between human and AI companions to facilitate the better design of AI companionship. All the participants' experimental data was stored on a local device with encryption.

\section{Limitation and Future Work}

We acknowledge several limitations in our study, which also suggest directions for future research. First, participants were recruited via the online social platform Red Book, which is popular among young adults for sharing personal experiences. This may introduce sampling bias. Second, our study primarily used semi-structured interviews. Incorporating methods like ethnography could provide deeper insights into the interaction patterns between humans and AI companions.

\section{Conclusion}

This paper explores the intricacies of human-AI intimate companionship within a Chinese context, offering insights into how individuals establish and maintain these relationships. Through qualitative research methods, including semi-structured interviews, we identified key themes in user motivations, interaction patterns, and the concerns associated with AI companions. These findings contribute to the broader discourse on the social and emotional dimensions of AI relationships, addressing both the opportunities and risks involved, such as over-immersion and privacy concerns. We underscore the importance of recognizing AI's role as a companion in modern society and its potential to fulfill emotional needs. We also calls for a cautious approach, emphasizing the need for more inclusive and ethical designs that consider users' diverse social and psychological experiences. Future research should aim to explore these relationships further, particularly focusing on long-term psychological impacts and the evolving dynamics as AI technologies continue to advance.

\bibliographystyle{ACM-Reference-Format}
\bibliography{sample-base}

\end{document}